\documentclass[12pt,manuscript]{aastex}

\slugcomment{{\em}}

\begin{document}

\title{The Formation of Disks in Elliptical Galaxies}

\author{Thorsten Naab \& Andreas Burkert}
\affil{Max-Planck-Institut f\"ur Astronomie, K\"onigstuhl 17,\\
       D-69117 Heidelberg, \\ Germany}

\begin{abstract}
We investigate detailed kinematical properties of simulated
collisionless merger remnants of disk galaxies with mass ratios of 1:1
and 3:1. The simulations were performed by direct summation using the
new special hardware device GRAPE-5. In agreement with observations,
the shape of the line-of-sight velocity distribution (LOSVD) is
Gaussian with small deviations. For most cases we find that the
retrograde wings of the LOSVD are steeper than the prograde ones. This
is in contradiction with observations which show broad retrograde and
steep prograde wings. This serious problem in the collisionless
formation scenario of massive elliptical galaxies can be solved if all
rotating ellipticals, even boxy ones, contain an additional stellar
disk component with $\propto 15\%$ of the total stellar mass and a
scale length of order the effective radius of the spheroid. We propose
that the progenitor galaxies of massive ellipticals must have
contained a significant amount of gas that did not condense into stars
during the merger process itself but formed an extended gaseous disk
before the star formation epoch. The heating source that prevented the
gas from forming stars early and the origin of the large specific
angular momentum required for the gas component to form an extended
disk are still unsolved problems.
\end{abstract}


\section{Introduction}
The merger hypothesis \citep{TT1972} assumes that elliptical galaxies
form by major mergers of spiral galaxies. This is supported by
observations of nearby merging systems that seem to evolve into
systems with properties comparable to elliptical galaxies
\citep{HY1999}. The gravitational interaction during the merger
process also triggers massive star formation and, probably, the
formation of massive central black holes and AGNs
\citep{SM1996,RSG1999}. On the theoretical side, numerous numerical 
simulations have shown that mergers lead to spheroidal systems with
global properties comparable to elliptical galaxies
\citep{BH1992,B1998,BS1997,NBH1999}. During the merging epoch the
galaxies are far from dynamical equilibrium. Gas, if added to the
simulations, condenses into stars by massive bursts of star formation
during the merging process or effectively looses angular momentum and
falls into the central parts of the merger remnant, probably feeding a
central black hole \citep{MH1996,S2000,B2000}. When the merger
remnants have settled into dynamical equilibrium, phase mixing and
violent relaxation have erased most information about the initial
conditions. However, as violent relaxation is incomplete, the stellar
phase space distribution of elliptical galaxies should still show fine
structure which affects the isophotal shapes and velocity
distributions. This fine structure provides fundamental insight into
the formation epoch of elliptical galaxies.  

Numerical simulations of collisionless mergers of disk galaxies
have confirmed this conclusion. The LOSVDs of merger remnants and
the orbit structure of their stars contain information about the
initial disk orientations \citep{HHS1996}. The isophotal shapes and
the dynamics of the merger remnants are primarily determined by their
initial mass ratios \citep{NBH1999,BB2000}.

In this letter we show that the LOSVDs of most elliptical galaxies
contain evidence for a second disk component with roughly 10\% to 20\%
the luminosity of the spheroid and a large scale radius which if is
order the effective radius of the spheroid. This disk might have
formed by late gas infall from gas-rich progenitors, followed by star
formation. Observationally, photometric and kinematic investigations of
elliptical galaxies have already indicated the existence of large
stellar disks of the same order of magnitude as predicted here,
especially in disky, rotating ellipticals \citep{RW1990,SBW1998}. We
demonstrate that this component is required in almost all ellipticals,
independent of whether they are disky or boxy in order to bring the
observations in agreement with the merger scenario of the formation of
early-type galaxies. 
 
\section{The merger models}

We use the method described by \citet{Her1993} to construct disk
galaxies in dynamical equilibrium. The system of units is:
gravitational constant G=1, exponential scale length of the larger
disk in the merger h=1 and mass of the larger disk $M_d=1$. 
The disks are exponential with an additional spherical, non-rotating
bulge with mass $M_b = 1/3$, a Hernquist density profile \citep{Her1990}
and a scale length $r_b=0.2h$. The system lives in a spherical
pseudo-isothermal halo with a mass $M_d=5.8$, cut-off radius $r_c=10h$
and core radius $\gamma=1h$. 

The N-body simulations were performed using a direct summation code
with the new special purpose hardware GRAPE-5 \citep{KFM2000}.
The 1:1 merger was calculated adopting in total 400000 particles with
each galaxy consisting of 60000 disk particles, 20000 disk particles
and 120000 halo particles. For the 3:1 merger the parameters of the
more massive galaxy were as described above. The low-mass galaxy
contained 1/3 the mass and number of particles of the larger galaxy, with
a disk  scale length of $h=\sqrt{1/3}$, as expected from the
Tully-Fisher relation. 

For both mass ratios, the galaxies approached each other on nearly
parabolic orbits with an initial separation of 30 length units
and a pericenter distance of 2 length units. In this letter we
focus on one corotating and one counterrotating geometry. We have
found that these two impact parameters are representative for a much
larger set of simulations that have been performed and that will be
discussed in a subsequent paper. The inclinations of the corotating
disk models (called model A for the equal mass merger and model C for
the 3:1 merger) relative to the orbital plane are $t_1= 30^{\circ}$
and $t_2 =-30^{\circ}$ with arguments of pericenter of $\omega_1 =
30^{\circ}$ and $ \omega_2 =-30^{\circ}$, respectively. The
counterrotating disks (called model B for 1:1 and model D for 3:1)
have  $t_1= 30^{\circ}$ and $t_2 = 150^{\circ}$, $\omega_1 =
30^{\circ}$ and $ \omega_2 =-30^{\circ}$, respectively.  In all
simulations the merger remnants were allowed to settle into dynamical
equilibrium for approximately  10 dynamical timescales after the
merger was complete. Then their equilibrium state was analysed. 

\section{LOSVD analysis of the merger remnants}

To measure the line-of-sight velocity distribution of a merger remnant
we shifted the densest region of every two-dimensional projection to
the origin. Then we placed a slit with a width of 0.2 unit lengths
along the apparent long axis of each projected remnant. Thereafter we
binned all particles falling within each grid cell in velocity along
the line-of-sight. The grid spacing was chosen to be 0.15 times the
projected half-mass radius. The width of the velocity bins was set to
a value of 0.2 for line-of-sight velocities $v_{\mathrm{los}}$ in the
range $ -4 \ge v_{\mathrm{los}} \ge 4$. This results in 80 velocity
bins over the whole velocity interval. Using the binned velocity data
we constructed line-of-sight velocity profiles (LOSVD) for each bin
along the grid. Subsequently we parametrized deviations from the
Gaussian shape of the velocity profile using Gauss-Hermite basis
functions \citep{vdMF1993,G1993,BB2000}. The kinematic parameters of
each profile ($\sigma_{\mathrm{fit}}$, $v_{\mathrm{fit}}$, $H_3$,
$H_4$) were then determined by least squares fitting. The large number
of simulated stellar particles ($>$ 100000) guaranteed that at least
2000 particles fall within each slit inside one effective radius.

\section{Comparison with observations}

The straight lines in Figure 1 indicate the observed local correlation
between $H_3$ and $v/\sigma$ for a sample of elliptical galaxies in
low density environments, published by \citet{BSG1994}. In all cases
$H_3$ and $v/\sigma$ have opposite signs. The data indicate that all
elliptical galaxies have LOSVDs with steep prograde wings and broad
retrograde wings.  \citet{MSB2000} have shown that this result also
holds for cluster ellipticals in Coma.

The local correlation between $H_3$ and $v/\sigma$ for the simulated
corotating (A,C) and counterrotating (B,D) 1:1 and 3:1 mergers,
respectively, is shown by the dots in Figure 1. Every remnant is
analysed as seen from 50 random viewing angles. For the corotating
mergers the correlation between $H_3$ and $v/\sigma$ is almost
opposite to the observed one. This is also reflected in a positive
effective value for $H_3$ (left plot in Figure 2) for almost all
projection angles, where $<H_3>$ is defined as the mean value between
0.25 and 0.75 effective radii \citep{BSG1994}. Here, a positive
$<H_3>$ corresponds to an anticorrelation between $H_3$ and
$v/\sigma$, such that $H_3 >0$ for $v/\sigma <0$ and vice versa. We
find that $<H_3>$ versus $v/\sigma_0$ does not follow the observed
correlation, indicated by the solid and dashed lines.  The profiles
have broad prograde wings and narrow retrograde wings. The
counterrotating 3:1 merger (D) in Fig. 1 shows a large spread around
zero but also does not agree with observations.  The only exception is
the equal mass merger of counter-rotating disks (B) which leads to a
very anisotropic elliptical with no signature of rotation.

Note, that the four simulations which we focus on in this paper have
been chosen because they are representative for a much larger set of
simulations with different orbital geometries. These four models have
been recalculated with high resolution for better statistics.

We conclude that collisionless mergers of disk galaxies in general
fail to explain the detailed kinematics of all observed elliptical
galaxies (also massive, boxy ones) that exhibit a significant amount
of rotation ($v/\sigma \ge 0.2$) inside one $r_{\mathrm{eff}}$.
Further evidence for a possible failure of the collisionless merger
picture comes from \citet{CNR2001} who showed that the kinematical
properties of very faint and fast rotating, disky elliptical galaxies
can not be explained by collisionless mergers of disk galaxies.

\section{Theoretical evidence for disks in elliptical galaxies?}

One plausible explanation for line-of-sight velocity distributions
with negative $H_3$ is the superposition of a spheroidal body with a
disk-like component \citep{BSG1994}.  With a simple experiment one can
test if the wrong correlations found in our remnants result from a
lack of such a component. We artificially added a thin (scale height
$\approx 0.05 r_{\mathrm{eff}}$), cold (velocity dispersion
perpendicular to the disk $\sigma_z = 0$), stellar disk with an
exponential surface density profile. The disk was placed in the plane
defined by the long and intermediate axes of the main stellar body,
rotating in the same direction as the main stellar body. The particles
of the disk were assumed to move around the center of the galaxy in
centrifugal equilibrium with the gravitational potential arising from
the total enclosed mass. No additional random motion was added. Under
these simple assumptions only two parameters remain free: the total
mass of the disk $M_d$ in units of the total luminous mass of the
remnant and its scale length $r_d$ in units of the projected
half-light radius $r_{\mathrm{eff}}$. The results for the corotating
3:1 remnant (C) are summarized in the Figures 3 and 4. This case is
representative for all our unequal mass merger remnants.  In general,
we find that disks with small masses or radii do not change the LOSVDs
of the stellar component. Disks with masses and radii in the region
indicated by the black dots in Figure 4, on the other hand, lead to a
significant change in the resulting line profile. The prograde wings
become steeper than the retrograde ones in very good agreement with
observations.  This effect is shown in Figure 3 for a disk with 15\%
of the spheroid mass and $r_d = 1.25 r_{\mathrm{eff}}$. The influence
on the $<H_3>$ versus $v/\sigma_0$ correlation and its dependence on
projection effects is shown on the right hand side of Figure 2. Now
the values fall in the observed regime.

If the additional disk becomes too massive, the absolute values for
$H_3$ are larger than observed and the surface brightness profiles
change from de Vaucouleurs to exponential profiles which is again not
in agreement with observations. We therefore conclude that the
existence of an additional stellar disk component with 10\% to 20\%
the luminosity of the spheroid and a scale length of the order of
$r_{\mathrm{eff}}$ can explain the observed correlation between $H_3$
and $v/\sigma$ in elliptical galaxies.

\section{Discussion and Conclusions}

Our simulations indicate that models of pure collisionless mergers in
general fail to reproduce LOSVDs of observed elliptical galaxies.  All
simulated profiles locally have the wrong sign of $H_3$ with respect
to $(v/\sigma)$, compared with the observations. A similar result has
been reported by \citet{BB2000}. They, however, did not investigate
projection effects. A detailed investigation of local properties of
the LOSVDs and projection effects shows that the presence of an
additional stellar disk component added to the stellar body after the
merger is complete could solve the problem.

One possible origin for such a disk is gas that must have been present
during the formation epoch of massive ellipticals and that formed a
disk after the major merger event was completed. In order to fit the
observed profiles, the amount of gas that settled into the equatorial
plane must have been significant. It is puzzling why the gas did not
turn into stars prior to its infall into the equatorial plane and how
the gas could keep its angular momentum required for the disk to have
a scale length equal to the effective radius of the spheroid.

\citet{BH1996} do find gaseous disk-like components in their equal
mass merger models. Tidal torques lead however to efficient angular
momentum loss in the gaseous component resulting in gas infall to the
center. Only 20\% of the initial gas mass (which was 10\% of the
initial disk mass) and by this less than 2\% of the stellar mass
settled into an extended disk-like component.  In this case the disk
would not be massive enough to change the LOSVD of the system
significantly. Equal mass mergers of very gas rich disks (more than
50\% gas) could, however, lead to the formation of more massive
disk-like components in the end.

First simulations by \citet{NB2001} have shown that for unequal mass
mergers of gas rich galaxies which lead to disky, fast rotating
ellipticals due to the stronger centrifugal support, a large fraction
of the gas settles into extended disks after the merger is complete,
if star formation is suppressed. This seems to be an attractive
scenario to explain the existence of large disks in disky elliptical
galaxies.  The origin of large stellar disks in massive boxy
ellipticals with a small but significant amount of rotation that form
presumably from equal-mass mergers is, however, still unclear.

\acknowledgments

We thank Ralf Bender, Hans-Walter Rix and Nicola Cretton for helpful
discussions.

\clearpage
\begin{figure}
\epsscale{0.5}
\plotone{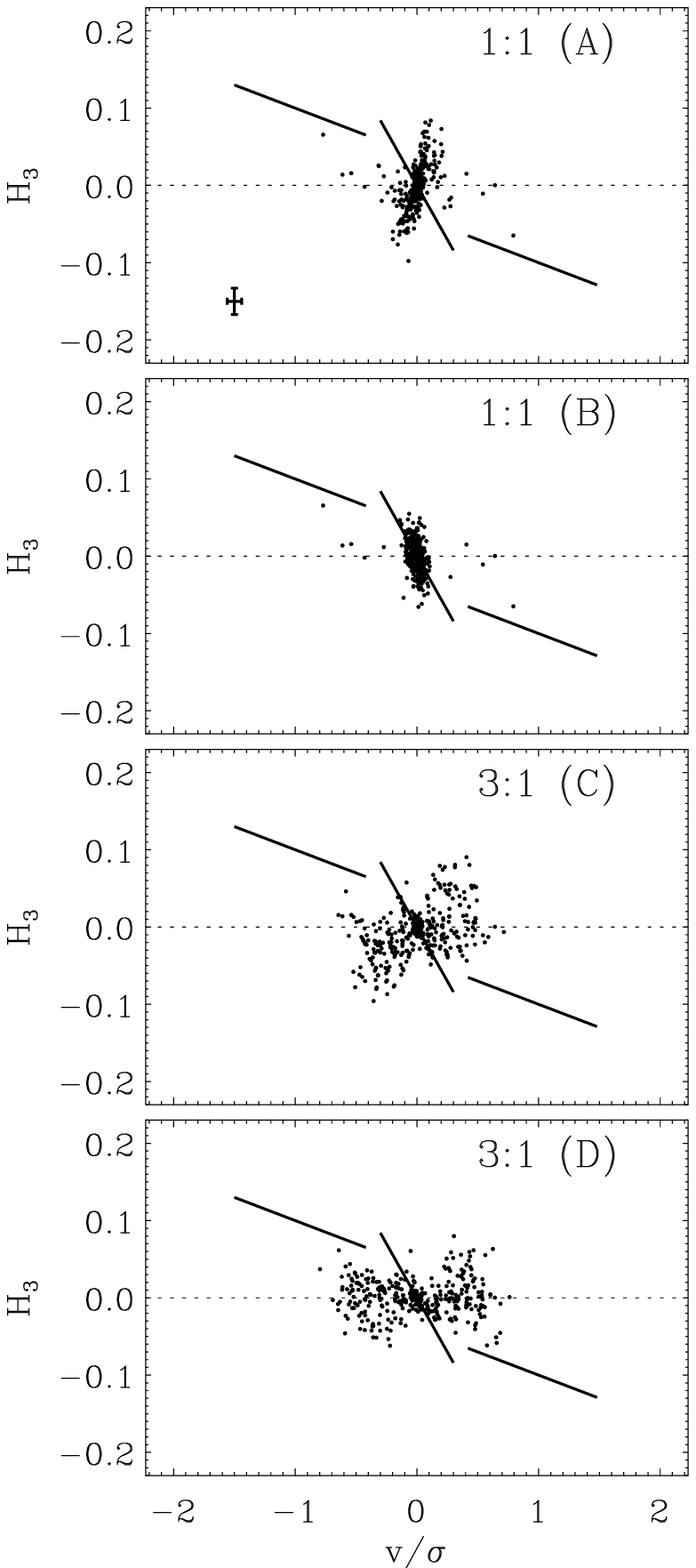}
\caption{Local correlation between $H_3$ and $v/\sigma$ for the 1:1
and 3:1 merger simulations. The models (A,C) and (B,D) have co -and
counterrotating geometries, respectively. The dots represent the data
for each model as seen from 50 random viewing angles. The error bar in
the upper panel shows the bootstrap error at 0.75
$r_{\mathrm{eff}}$. The observed correlation from \citet{BSG1994} is
tentatively indicated by the straight lines. For the complete data set see
their Figure 15.\label{fig1}}  
\end{figure}

\clearpage

\begin{figure}
\epsscale{1.0}
\plotone{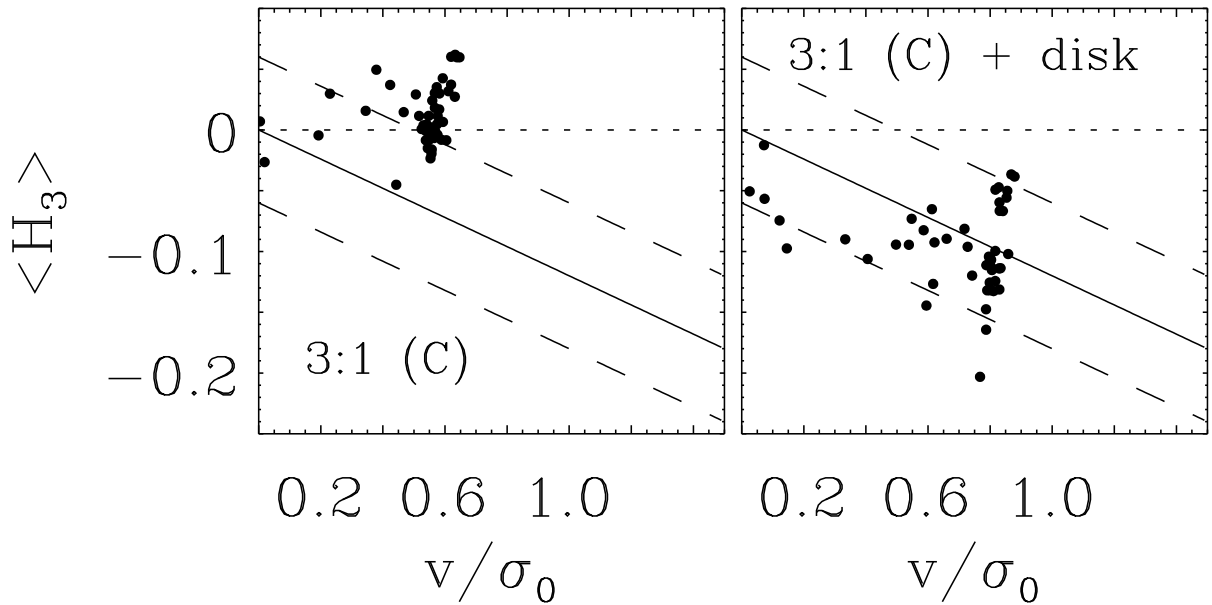}
\caption{Left:Effective $H_3$ versus $v/\sigma_0$
(rotational velocity at one $r_{\mathrm{eff}}$ over central velocity
dispersion) for 50 random projection of model (C). Right: Same for
model (C) with an additional disk as described in the text. The
straight line is the mean correlation observed by \citet{BSG1994}. The
parallel long-dashed lines give an indication of the observed
spread. The short dashed line is $<H_3>$ = 0.\label{fig2}}
\end{figure}
\clearpage

\begin{figure}
\plotone{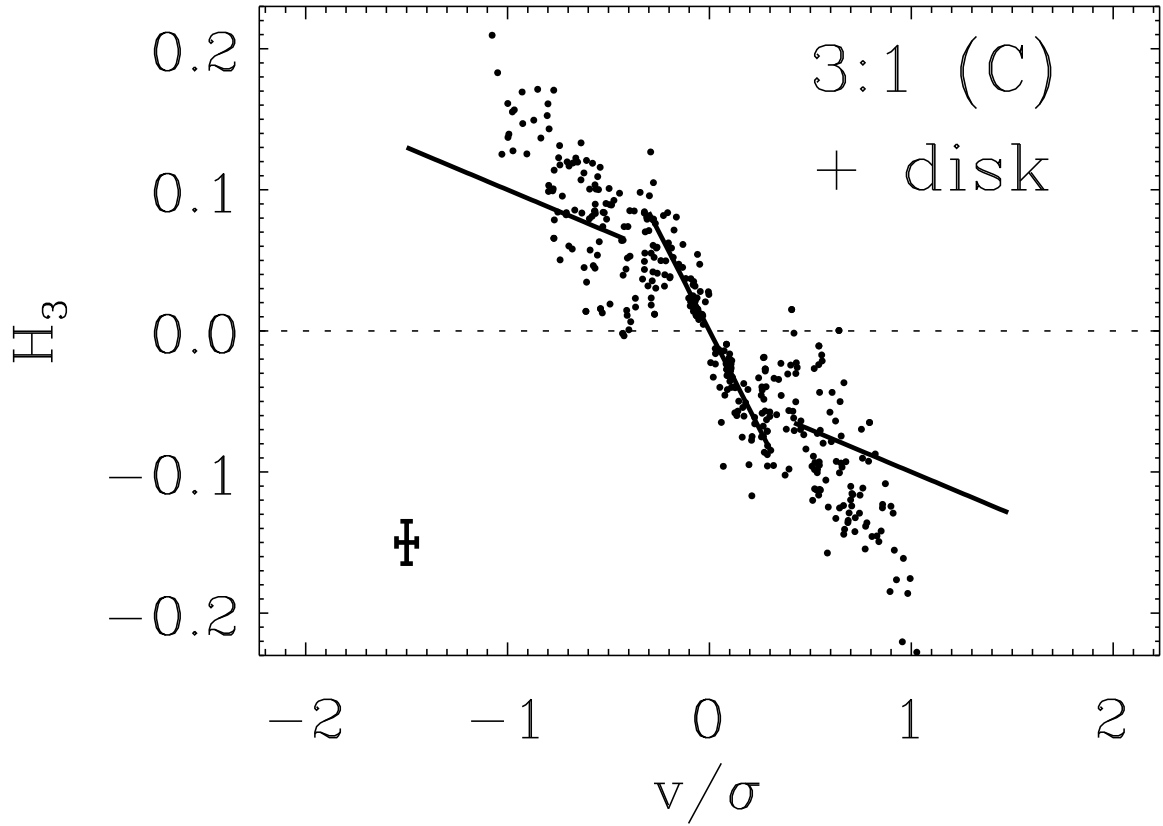}
\caption{Local correlation between $H_3$ and $v/\sigma$ for model (C)
with an additional thin stellar exponential disk with a scale length
of $1.25 r_{\mathrm{eff}}$. A mass of $15\%$ of the spheroid is
assumed. The error is derived in the same way as in
Figure 1. \label{fig3}}
\end{figure}

\clearpage
\begin{figure}
\plotone{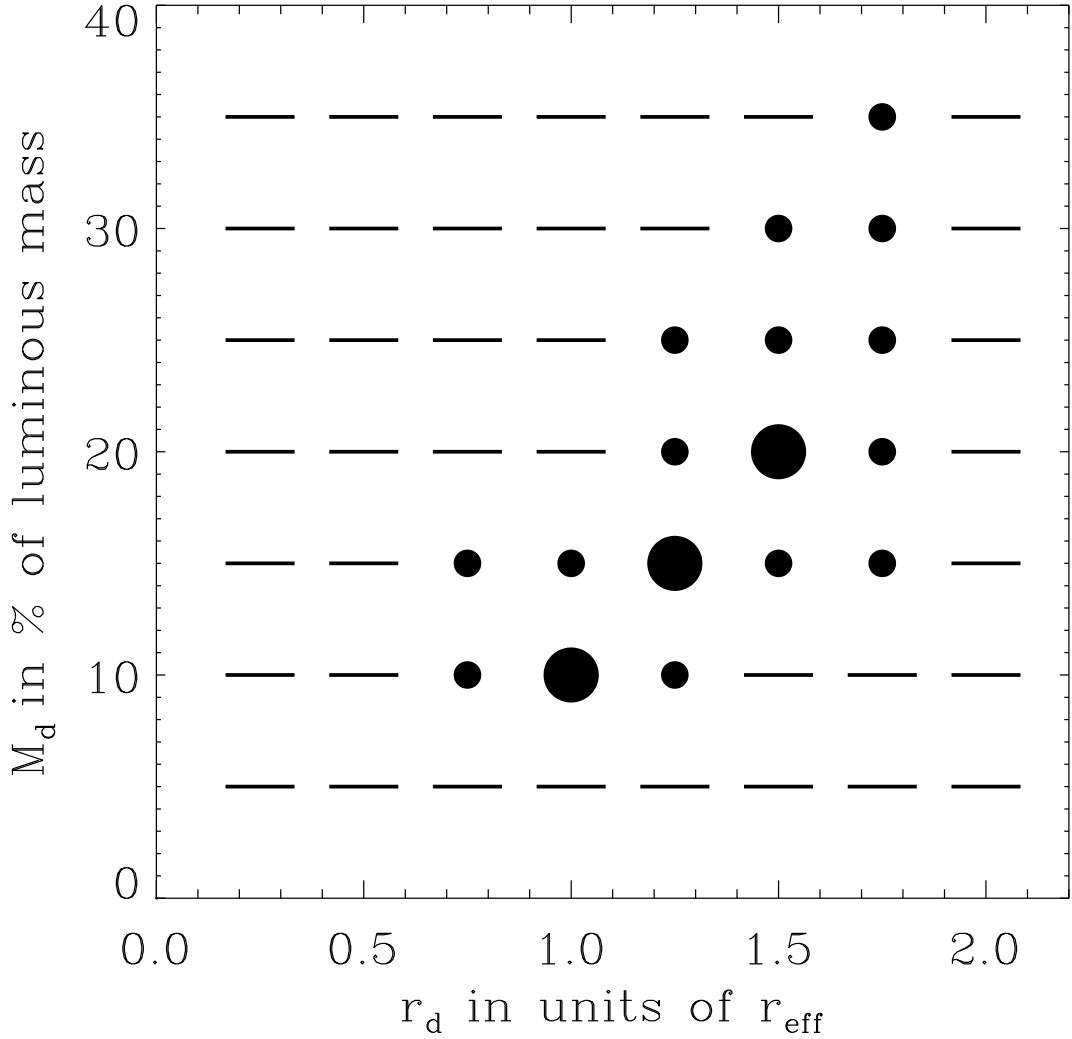}
\caption{Mass $M_d$ versus scale radius $r_d$ for a disk
added to the merger remnant (C). Combinations that are able to
reproduce (big dot), almost reproduce (small dot) or fail to reproduce
(minus sign) the observed correlations are shown. \label{fig4}}
\end{figure}
\end{document}